\documentclass{ws-procs9x6}
\usepackage{revsymb}

\def\vsk#1{\noalign{\vskip#1 cm}}

\begin{document}
\begin{flushright}
{\small OCHA-PP-270, YITP-07-09, VPI-IPNAS-07-02}
\end{flushright}

\title{The effect of Topcolor Assisted Technicolor, and other models,\\
on Neutrino Oscillation}
\author{Minako~Honda${}^1$, Yee~Kao$^2$, Naotoshi~Okamura${}^3$, Alexey~Pronin${}^2$, and Tatsu~Takeuchi${}^2$\footnote{Presenting Author}}
\address{
${}^1$Physics Department, Ochanomizu Women's University, Tokyo 112-8610, Japan\\
${}^2$Physics Department, Virginia Tech, Blacksburg VA 24061, USA\\
${}^3$Yukawa Institute for Theoretical Physics, Kyoto University, Kyoto 606-8502, Japan}
\date{\today}

\begin{abstract}
\noindent
New physics beyond the Standard Model
can lead to extra matter effects on neutrino oscillation
if the new interactions distinguish among the three flavors of neutrino.
In Ref.~\refcite{HOT}, we argued that a long-baseline neutrino oscillation experiment 
in which the Fermilab-NUMI beam in its high-energy mode \cite{NUMI}
is aimed at the planned Hyper-Kamiokande detector \cite{HyperK}
would be capable of constraining the size of those extra matter effects, provided the vacuum value of $\sin^2 2\theta_{23}$ is not too close to one.
In this talk, we discuss how such a constraint would translate into limits on
the coupling constants and masses of new particles 
in models such as topcolor assisted technicolor \cite{TopTechni}.
\end{abstract}

\bodymatter
\section{Introduction}

When considering matter effects on neutrino oscillation, it is customary to consider only the $W$-exchange interaction of the $\nu_e$ with the electrons in matter.
However, if new interactions beyond the Standard Model (SM) that distinguish among the three generations of neutrinos exist, they can lead to extra matter effects via radiative corrections to the
$Z\nu\nu$ vertex which effectively violate neutral current universality, 
or via the direct exchange of new particles between the neutrinos and matter particles.  

For instance, topcolor assisted technicolor\cite{TopTechni} 
treats the third generation differently from the first two and the
$Z'$ in this class of models couples more strongly to the $\nu_\tau$ than
to the $\nu_e$ or $\nu_\mu$.
In Extended Technicolor (ETC) Models, such as that of Appelquist, Piai, and Shrock\cite{APS}, the neutral technimesons, which mix with the $Z$, couple to
different generation fermions differently, distinguishing among
$\nu_e$, $\nu_\mu$, and $\nu_\tau$.  
The diagonal ETC gauge bosons 
also couple to the different generations
differently, as well as the large variety of leptoquark states in the model.
Flavor distinguishing matter effects from diagonal ETC and leptoquarks are induced by ETC gauge boson mixing.

The effective Hamiltonian that governs neutrino oscillations
in the presence of neutral-current lepton universality violation, or new physics
that couples to the different generations differently, is given by~\cite{HOT}
\begin{equation}
H = 
\tilde{U}
\left[ \begin{array}{ccc} \lambda_1 & 0 & 0 \\
                          0 & \lambda_2 & 0 \\
                          0 & 0 & \lambda_3
       \end{array}
\right]
\tilde{U}^\dagger
= U
\left[ \begin{array}{ccc} 0 & 0 & 0 \\
                          0 & \delta m^2_{21} & 0 \\
                          0 & 0 & \delta m^2_{31}
       \end{array}
\right]
U^\dagger +
\left[ \begin{array}{ccc} a & 0 & 0 \\
                          0 & 0 & 0 \\
                          0 & 0 & 0 
       \end{array}
\right] +
\left[ \begin{array}{ccc} b_e & 0 & 0 \\
                          0 & b_\mu & 0 \\
                          0 & 0 & b_\tau 
       \end{array}
\right] \;,
\label{Hdef}
\end{equation}
where $U$ is the MNS matrix,
\begin{equation}
a=2E V_{CC}\;,\qquad V_{CC} = \sqrt{2} G_F N_e = N_e \dfrac{g^2}{4 M_W^2}\;,
\end{equation}
is the usual matter effect due to $W$-exchange
between $\nu_e$ and the electrons, and 
$b_e$, $b_\mu$, $b_\tau$
are the extra matter effects which we assume to be non-equal. 
%
%
We define the parameter $\xi$ as
\begin{equation}
\dfrac{b_\tau - b_\mu}{a} = \xi \;.
\end{equation}
Then, the effective Hamiltonian can be rewritten as
\begin{equation}
H = 
\tilde{U}
\left[ \begin{array}{ccc} \lambda_1 & 0 & 0 \\
                          0 & \lambda_2 & 0 \\
                          0 & 0 & \lambda_3
       \end{array}
\right]
\tilde{U}^\dagger
= U
\left[ \begin{array}{ccc} 0 & 0 & 0 \\
                          0 & \delta m^2_{21} & 0 \\
                          0 & 0 & \delta m^2_{31}
       \end{array}
\right]
U^\dagger + a
\left[ \begin{array}{ccc} 1 & 0 & 0 \\
                          0 & -\xi/2 & 0 \\
                          0 & 0 & +\xi/2 
       \end{array}
\right] \,,
\label{Hdef2}
\end{equation}
where we have absorbed the extra $b$-terms in the $(1,1)$ element into $a$.

The extra $\xi$-dependent contribution in Eq.~(\ref{Hdef2}) 
can manifest itself when $a>|\delta m^2_{31}|$
(\textit{i.e.} $E\agt 10\,\mathrm{GeV}$ for typical matter densities in the Earth)
in the $\nu_\mu$ and $\bar{\nu}_\mu$ survival probabilities as \cite{HOT}
\begin{eqnarray}
{P}(\nu_\mu\rightarrow\nu_\mu) 
& \approx & 1-\sin^2\left(2\theta_{23} - \frac{a\xi}{\delta m^2_{31}}\right)\sin^2\dfrac{{\Delta}}{2}\;, \cr
{P}(\bar{\nu}_\mu \rightarrow \bar{\nu}_\mu)
& \approx & 1-\sin^2\left(2\theta_{23} + \frac{a\xi}{\delta m^2_{31}}\right)\sin^2\dfrac{{\Delta}}{2}\;,
\end{eqnarray}
where 
\begin{equation}
\Delta \approx \Delta_{31} c_{13}^2 - \Delta_{21} c_{12}^2\;,\qquad
\Delta_{ij} = \dfrac{\delta m^2_{ij}}{2E}L\;,\qquad
c_{ij} = \cos\theta_{ij}\;,
\end{equation}
and the CP violating phase $\delta$ has been set to zero.
As is evident from these expressions, the small shift due to $\xi$ will be invisible if
the value of $\sin^2 2\theta_{23}$ is too close to one.
However, if the value of $\sin^2 2\theta_{23}$ is as low as 
$\sin^2 2\theta_{23}=0.92$ (the current 90\% lower bound), and if
$\xi$ is as large as $\xi =0.025$ (the central value of from CHARM/CHARM~II \cite{CHARM}),
then the shift in the survival probability at the first oscillation dip can be as large as $\sim 40\%$.
If the Fermilab-NUMI beam in its high-energy mode \cite{NUMI} 
were aimed at a declination angle of $46^\circ$ toward the planned 
Hyper-Kamiokande detector \cite{HyperK} in Kamioka, Japan
(baseline 9120~km), such a shift would be visible after just one year of data taking, assuming
a Mega-ton fiducial volume and 100\% efficiency.
The absence of any shift after 5 years of data taking would constrain $\xi$ to \cite{HOT}
\begin{equation}
|\xi| \le \xi_0 \equiv 0.005\;,
\label{xi_bound}
\end{equation}
at the 99\% confidence level.

In the following, we look at how this potential limit on $\xi$ would translate into constraints on the $Z'$ in topcolor assisted technicolor, and various
types of leptoquarks.
A more comprehensive analysis will be presented in Ref.~\refcite{HKOPT}.

\section{Topcolor Assisted Technicolor}

Though there are several different versions of 
topcolor assisted technicolor\cite{TopTechni}, 
we consider here the simplest in which the quarks and leptons transform under the gauge group
\begin{equation}
SU(3)_s \times SU(3)_w \times 
 U(1)_s \times  U(1)_w \times SU(2)_L
\end{equation}
with coupling constants $g_{3s}$, $g_{3w}$, $g_{1s}$, $g_{1w}$,
and $g$, respectively.
It is assumed that $g_{3s} \gg g_{3w}$ and $g_{1s} \gg g_{1w}$.
$SU(2)_L$ is the usual weak-isospin gauge group of the SM.
The first and second generation fermions are assumed to be charged only under
$SU(3)_w\times SU(2)_L\times U(1)_w$, 
while the third generation fermions are assumed to be charged only under
$SU(3)_s\times SU(2)_L\times U(1)_s$. 
The $U(1)$ charges for both cases are set equal to the SM hypercharge.
At scale $\Lambda\sim 1$~TeV, technicolor, which is included in the model to generate the $W$ and $Z$ masses, is assumed to become strong and generate a condensate (of something which is left unspecified)
which breaks the two $SU(3)$'s and the two $U(1)$'s to their diagonal
subgroups:
\begin{equation}
SU(3)_s \times SU(3)_w \rightarrow SU(3)_c\;,\qquad
U(1)_s \times U(1)_w \rightarrow U(1)_Y\;,
\end{equation}
which we identify
with the usual SM color and hypercharge groups.
The massless unbroken U(1) gauge boson $B_\mu$ and 
the massive broken U(1) gauge boson $Z'_\mu$
are related to the original $U(1)_s\times U(1)_w$ gauge fields
$Y_{s\mu}$ and $Y_{w\mu}$ by
\begin{eqnarray}
Z'_\mu & = & Y_{s\mu} \cos\theta_1 - Y_{w\mu} \sin\theta_1 \cr
B_\mu  & = & Y_{s\mu} \sin\theta_1 + Y_{w\mu} \cos\theta_1
\end{eqnarray}
where
\begin{equation} 
\tan\theta_1 = \frac{g_{1w}}{g_{1s}}\;.
\label{theta1}     
\end{equation}
The currents to which the $B_\mu$ and $Z'_\mu$ couple to are:
\begin{equation}
g_{1s}J_{1s}^\mu Y_{s\mu} + g_{1w}J_{1w}^\mu Y_{w\mu}
= g'\left( \cot\theta_1 J_{1s}^\mu - \tan\theta_1 J_{1w}^\mu
     \right) Z'_\mu
+ g'\left( J_{1s}^\mu + J_{1w}^\mu \right) B_\mu\;,
\label{ZprimeCoupling}
\end{equation}
where
\begin{equation} 
\frac{1}{g^{\prime 2}} = \frac{1}{g_{1s}^2} + \frac{1}{g_{1w}^2}\;.  
\end{equation}
The current $J_Y^\mu = J_{1s}^\mu + J_{1w}^\mu$
is the SM hypercharge current, and $g'$ is the SM hypercharge
coupling constant.  

The exchange of the $Z'$ leads to the current-current interaction
\begin{equation}
\dfrac{g^{\prime 2}}{2 M_{Z'}^2}
\left( \cot\theta_1 J_{1s} - \tan\theta_1 J_{1w} \right)
\left( \cot\theta_1 J_{1s} - \tan\theta_1 J_{1w} \right)
\;,
\label{Current_Current}
\end{equation} 
the $J_{1s}J_{1s}$ part of which does not contribute to neutrino oscillations on the Earth, 
while the $J_{1w}J_{1w}$ part is suppressed relative to the $J_{1w}J_{1s}$ part by a factor of $\tan^2\theta_1 \ll 1$.
Therefore, we only need to consider the $J_{1s}J_{1w}$ interaction which only affects the propagation of $\nu_\tau$.
The effective potential felt by $\nu_\tau$ due to this interaction is \cite{HKOPT}
\begin{equation}
V_{\nu_\tau} 
\;=\; \dfrac{N_n}{8}\dfrac{g^{\prime 2}}{M_{Z'}^2} 
\;\approx\; \dfrac{N_e}{8}\dfrac{g^{\prime 2}}{M_{Z'}^2} \;,
\end{equation}
and the effective $\xi$ is
\begin{equation}
\xi_{TT} 
= \frac{V_{\nu_\tau} - V_{\nu_\mu}}{V_{CC}}
= \frac{1}{2}\,\dfrac{(g'/M_{Z'})^2}{(g/M_W)^2}
= \frac{1}{2}\tan^2\theta_W\,\dfrac{M_W^2}{M_{Z'}^2}
= \frac{1}{2}\sin^2\theta_W\,\dfrac{M_Z^2}{M_{Z'}^2} 
\;.
\end{equation}
The limit $|\xi_{TT}|\le \xi_{0} = 0.005$ then translates into:
\begin{equation}
M_{Z^{\prime}} \ge 
M_Z \sqrt{ \dfrac{\sin^2\theta_W}{2\xi_0} }
\approx 440\,\mathrm{GeV}\;.
\end{equation}
Unfortunately,
this potential limit from the measurement of $\xi$ 
is weaker than what is already available from precision electroweak data 
\cite{TopTechni_limits}, and from direct searches for $p\bar{p}\rightarrow Z'\rightarrow \tau^+\tau^-$ at CDF\cite{Yao:2006px,Acosta:2005ij}.

\section{Generation Non-diagonal Leptoquarks}

The interactions of leptoquarks with ordinary matter 
can be described in a model-independent fashion 
by an effective low-energy Lagrangian as discussed in 
Ref.~\refcite{leptoquarks}.
Assuming the fermionic content of the SM, the most general
dimensionless $SU(3)_C\times SU(2)_L\times U(1)_Y$ invariant couplings of
scalar and vector leptoquarks satisfying baryon and lepton number 
conservation are given by:
\begin{equation}
{\cal L} = {\cal L}_{F=2} + {\cal L}_{F=0}\;,
\end{equation}
where
\begin{eqnarray}
\lefteqn{
{\cal L}_{F=2} 
\;=\;
\Bigl[
g_{1L}^{ij}(\overline{u_{iL}^c}e_{jL}^{\phantom{c}} - \overline{d_{iL}^c}\nu_{jL}^{\phantom{c}} )
+g_{1R}^{ij}(\overline{u_{iR}^c}e_{jR}^{\phantom{c}})
\Bigr] S_{1}^0
}
\cr
&+&\Bigl[
g_{2L}^{ij}(\overline{d_{iR}^c}\gamma^\mu e_{jL}^{\phantom{c}})
+g_{2R}^{ij}(\overline{d_{iL}^c}\gamma^\mu e_{jR}^{\phantom{c}}) 
\Bigr] V_{2\mu}^+
\cr
&+&\Bigl[
 g_{2L}^{ij}(\overline{d_{iR}^c}\gamma^\mu \nu_{jL}^{\phantom{c}})
+g_{2R}^{ij}(\overline{u_{iL}^c}\gamma^\mu e_{jR}^{\phantom{c}}) 
\Bigr] V_{2\mu}^-
\cr
&+&\tilde{g}_{2L}^{ij}\Bigl[
(\overline{u_{iR}^c}\gamma^\mu e_{jL}^{\phantom{c}}) \tilde{V}_{2\mu}^+
+ (\overline{u_{iR}^c}\gamma^\mu \nu_{jL}^{\phantom{c}}) \tilde{V}_{2\mu}^-
\Bigr]
\cr
&+&g_{3L}^{ij}\Bigl[-\sqrt{2}(\overline{d_{iL}^c}e_{jL}^{\phantom{c}}) S_{3}^+
             -(\overline{u_{iL}^c}e_{jL}^{\phantom{c}}+\overline{d_{iL}^c}\nu_{jL}^{\phantom{c}}) S_{3}^0
             +\sqrt{2}(\overline{u_{iL}^c}\nu_{jL}^{\phantom{c}}) S_{3}^- \Bigr] \;,
\\  \vsk{0.3}
\lefteqn{
{\cal L}_{F=0} 
\;=\;
\Bigl[
h_{2L}^{ij}(\overline{u_{iR}}e_{jL})+h_{2R}^{ij}(\overline{u_{iL}}e_{jR})
\Bigr] S_{2}^+
}
\cr
&+&\Bigl[
h_{2L}^{ij}(\overline{u_{iR}}\nu_{jL})-h_{2R}^{ij}(\overline{d_{iL}}e_{jR})
\Bigr] S_{2}^-
\cr
&+&\tilde{h}_{2L}^{ij} \Bigl[
(\overline{d_{iR}}e_{jL})\tilde{S}_{2}^+ + (\overline{d_{iR}}\nu_{jL})\tilde{S}_{2}^-
\Bigr] 
\cr
&+&\Bigl[
h_{1L}^{ij}(\overline{u_{iL}}\gamma^\mu \nu_{jL} +
       \overline{d_{iL}}\gamma^\mu e_{jL})
+h_{1R}^{ij}(\overline{d_{iR}}\gamma^\mu e_{jR}) \Bigr]V_{1\mu}^0
\cr
&+&h_{3L}^{ij}\Bigl[\sqrt{2}(\overline{u_{iL}}\gamma^\mu e_{jL}) V_{3\mu}^+
             +(\overline{u_{iL}}\gamma^\mu \nu_{jL} 
               -\overline{d_{iL}}\gamma^\mu e_{jL}) V_{3\mu}^0
             +\sqrt{2}(\overline{d_{iL}}\gamma^\mu \nu_{jL}) V_{3\mu}^- \Bigr]& \cr
&&
\end{eqnarray}
%
Here, the scalar and vector leptoquark fields are denoted by $S$ and $V$, 
their subscripts indicating the dimension of their $SU(2)_L$ representation,
and the superscripts indicating the sign of the weak-isospin of each component.
We allow for generation non-diagonal couplings with
the indices $i$ and $j$ indicating the quark and lepton generation numbers, 
respectively.
The subscript $L$ or $R$ on the coupling constants indicate 
the chirality of the lepton involved in the interaction.
For simplicity, color indices have been suppressed.
The leptoquarks $S_1, \vec{S}_3, V_2, \tilde{V}_2$ carry 
fermion number $F=3B+L=-2$, while
the leptoquarks $S_2, \tilde{S}_2, V_1, \vec{V}_3$ have $F=0$.
The interactions that affect neutrino oscillation are those with
$(ij)=(12)$ or $(13)$.

\begin{center}
\begin{table}[t] 
\tbl{Constraints on the leptoquark couplings with all the leptoquark masses set to 300~GeV. 
To obtain the bounds for a different leptoquark mass $M_{LQ}$, simply
rescale these numbers with the factor $(M_{LQ}/300\text{ GeV})^2$.}
{\begin{tabular}{|c||c|c|c|} 
\hline
$\;\; LQ \;\;$ 
& $\;\;\; C_{LQ} \;\;\;$ 
& $\quad\qquad\delta\lambda_{LQ}^2\qquad\quad$ 
& \ upper bound from $|\xi|\le\xi_0$\ \\
\hline\hline
$S_1$  
& $+3$  
& $ |g_{1L}^{12}|^2-|g_{1L}^{13}|^2$ 
& $0.01\phantom{0}$ \\
& & & \\
\hline
$\vec S_3$       
& $+9$                  
&  $|g_{3L}^{12}|^2-|g_{3L}^{13}|^2$  
& $0.003$ \\
\hline
$S_2$         
& $-3$                  
&  $|h_{2L}^{12}|^2-|h_{2L}^{13}|^2$  
& $0.01\phantom{0}$ \\
\hline
$\tilde{S}_2$   
& $-3$                  
& $|\tilde{h}{}_{2L}^{12}|^2-|\tilde{h}{}_{2L}^{13}|^2$  
& $0.01\phantom{0}$ \\
\hline
$V_2$          
& $+6$                  
& $|g_{2L}^{12}|^2-|g_{2L}^{13}|^2$  
& $0.005$ \\
\hline
$\tilde{V}_2$   
& $+6$                  
& $|\tilde{g}{}_{2L}^{12}|^2-|\tilde{g}{}_{2L}^{13}|^2$  
& $0.005$ \\
\hline
$V_1$            
& $-6$                  
& $|h_{1L}^{12}|^2-|h_{1L}^{13}|^2$ 
& $0.005$ \\
\hline
$\vec V_3$       
& $-18$                 
& $|h_{3L}^{12}|^2-|h_{3L}^{13}|^2$  
& $0.002$ \\
\hline
\end{tabular}}
\label{tab3}
\end{table}
\end{center}

It is straightforward to calculate 
the effective potentials due to the exchange of these leptoquarks, as well
as the effective values of $\xi$\cite{HKOPT}.
Assuming a common mass for leptoquarks in the same $SU(2)_L$ weak-isospin multiplet, the effective $\xi$ due to the exchange of any particular type of leptoquark can be written in the form
\begin{equation} \label{xi leptoquark}
\xi_{LQ}=C_{LQ}\;\frac{\;\delta\lambda_{LQ}^2/M_{LQ}^2\;}{g^2/M_W^2}=
\frac{C_{LQ}}{4\sqrt{2} G_F}\left(\frac{\delta\lambda^2_{LQ}}{M^2_{LQ}}\right)
\;.
\end{equation}
Here, $C_{LQ}$ is a constant prefactor, and $\delta\lambda_{LQ}^2$ represents
\begin{equation}
\delta\lambda_{LQ}^2 = |\lambda_{LQ}^{12}|^2 - |\lambda_{LQ}^{13}|^2\;,
\end{equation}
where $\lambda_{LQ}^{ij}$ is a generic coupling constant.
The values of $C_{LQ}$ and $\delta\lambda_{LQ}^2$ for the different types of leptoquark are listed in \tref{tab3}.
The constraint $|\xi_{LQ}|\le\xi_0$ translates into:
\begin{equation} 
\label{MLQbound}
M_{LQ} \ge 
\sqrt{\dfrac{|C_{LQ}| |\delta\lambda_{LQ}^2|}{4\sqrt{2}G_F\,\xi_0}}
\approx \sqrt{|C_{LQ}| |\delta\lambda_{LQ}^2|}\times (1700\,\mathrm{GeV}) \;.
\end{equation}
Alternatively, one can fix the leptoquark mass and obtain upper bounds on the leptoquark couplings:
\begin{equation}
\left|\delta\lambda^2_{LQ}\right|
\;\le\; \left(\dfrac{4\sqrt{2}G_F\,\xi_0}{C_{LQ}}\right) M^2_{LQ}
\;\approx\; \dfrac{\;0.03\;}{C_{LQ}}\left(\dfrac{M_{LQ}}{300\,\mathrm{GeV}}\right)^2\;.
\end{equation}
The values when $M_{LQ}=300\,\mathrm{GeV}$ are listed in the rightmost column of
\tref{tab3}.
Thought it is often stated that 
generation non-diagonal couplings of leptoquarks are
strongly constrained by the absence of flavor changing neutral currents,
it is only the products of the $(ij)=(12)$ and $(13)$ couplings 
with other couplings that are constrained\cite{davidson}. 
The limits on the individual couplings can be improved considerably.
The current leptoquark mass bounds from direct searches at 
the Tevatron, LEP, and 
HERA are in the 200$\sim$300~GeV range assuming 
generation diagonal couplings set equal to $\sqrt{4\pi\alpha}$.
At the LHC, leptoquarks, if they exist, can be expected to be pair-produced copiously through gluon-gluon fusion. 
The expected sensitivity is up to about 1.5~TeV\cite{LHC-LQ}. 
Depending on the value assumed for $\delta\lambda^2_{LQ}$, 
the bound from \eref{MLQbound} can be competitive.

\section*{Acknowledgments}

We would like to thank Drs. Andrew Akeroyd, Mayumi Aoki, Masafumi Kurachi,
Robert Shrock, and Hiroaki Sugiyama for helpful discussions.
This research was supported in part by the U.S. Department of Energy, 
grant DE--FG05--92ER40709, Task A (Kao, Pronin, and Takeuchi).

\bibliographystyle{ws-procs9x6}

\end{document}